\documentstyle[preprint,aps]{revtex}
\def\to{\rightarrow}
\begin{document}
\draft
\preprint{\vbox{\hbox{IFT--P.021/96}\hbox{hep-ph/9607241}}}
\title{Probing Higgs Couplings in \boldmath{$e^+ e^- \to
\gamma\gamma\gamma$}. } 
\author{F.\ de Campos, S.\ M.\ Lietti, S.\ F.\ Novaes and R.\ Rosenfeld}
\address{Instituto de F\'{\i}sica Te\'orica, 
Universidade  Estadual Paulista, \\  
Rua Pamplona 145, CEP 01405-900 S\~ao Paulo, Brazil.}
\date{\today}
\maketitle
\widetext
\begin{abstract}
We investigate the existence of anomalous Higgs boson couplings,
$H\gamma\gamma$ and $HZ\gamma$, through the analysis of the
process  $e^+ e^- \to \gamma\gamma\gamma$ at LEP2 energies. We
suggest some kinematical cuts to improve the signal to background
ratio and determine the capability of LEP2 to impose bounds on
those couplings by looking for a Higgs boson signal in this
reaction.
\end{abstract}


\section{Introduction}
\label{sec:int}

The predictions of Standard Model (SM) for the structure of the
fermion--vector boson couplings have been exhaustively tested in
the last few years. In particular, the recent data of LEP1 at the
$Z$--pole \cite{hep:conf} have confirmed with a unprecedented
degree of precision the properties of the neutral weak boson and
its vector and axial couplings with the different fermion
flavors. Nevertheless, we do not have the same level of
confidence on  other sectors of the SM, like the self--couplings
among the vector bosons and the Higgs boson couplings with
fermions and bosons. The determination of these interactions can
either confirm the non--abelian gauge structure of the theory and
the mechanism of the spontaneous breaking of the electroweak
symmetry or provide some hint about the existence of new physics
beyond the SM.

A convenient way  to parameterize possible deviations of the SM
predictions is through the effective theory approach
\cite{effective}. In this scenario, we assume that the existence
of new physics, associated to a high--energy scale $\Lambda$, can
manifest itself at low energy via quantum corrections, where the
heavy degrees of freedom are integrated out.  These effects are
then described by effective operators involving the spectrum of
particles belonging to the low--energy theory, {\it i.e.\/} the
usual fermions and bosons.

In the linear representation, a general dimension six effective
Lagrangian can be written as,
\begin{equation}
{\cal L}_{\text{eff}} = \sum_n \frac{f_n}{\Lambda^2} {\cal O}_n
\label{l:eff}
\end{equation}
where the operators ${\cal O}_n$ involve simultaneously both
vector boson and Higgs boson fields which share the same
coefficients $f_n$. Therefore, the study of anomalous Higgs boson
couplings can be an important tool to investigate the effect of
new physics and concomitantly furnish information about the
self--coupling of the vector bosons. In particular, anomalous
$H\gamma\gamma$ and $HZ\gamma$ couplings have already been
considered in $Z$ and Higgs boson decays \cite{Hagiwara2}, in
$e^+ e^-$ collisions \cite{Hagiwara2,epem} and at $\gamma\gamma$
colliders \cite{gg}. 

In this letter, we make an exhaustive analysis of the anomalous
Higgs boson contribution to the reaction  $e^+ e^- \to \gamma
\gamma \gamma$ at LEP2 in order to extract information about
possible anomalous $H\gamma\gamma$ and $HZ\gamma$ couplings. This
process is an ideal place to look for deviations from the SM
since it only involves well known purely QED contributions at
tree level \cite{3gamma}. Our Monte Carlo analysis of the
contribution $e^+ e^- \to \gamma, \; Z \to \gamma H (\to \gamma
\gamma)$ includes all the irreducible QED background and the
respective interferences. After detailed study of the signal and
background distributions, we find optimum cuts to maximize the
signal to background ratio. We show how to use energy and
invariant mass spectra of the final state photons in order to
identify the presence of a Higgs boson and extract information
about its couplings. Finally, we compare the bounds on the
anomalous couplings that could be provided by this reaction with
the present direct information on the triple vector boson
coupling.

\section{Anomalous Higgs Couplings and $\lowercase{e}^+
\lowercase{e}^- \to \gamma \gamma \gamma$}  
\label{sec:eff}

A convenient basis of dimension six operators that parametrizes
the deviations of the SM predictions in the bosonic sector was
constructed by Hagiwara {\it et al.} \cite{dim6:zep}. They
require the effective Lagrangian to be invariant under the local
$SU(2)_L \times U(1)_Y$ symmetry and to be C and P even. Of the
eleven operators ${\cal O}_{n}$ that form this basis and  induce
the Lagrangian (\ref{l:eff}), five affect the Higgs boson
interaction \cite{Hagiwara2,dim6:zep}, 
\begin{eqnarray}
{\cal O}_{BW} & = &  \Phi^{\dagger} \hat{B}_{\mu \nu} 
\hat{W}^{\mu \nu} \Phi \; , \nonumber \\ 
{\cal O}_{WW} & = & \Phi^{\dagger} \hat{W}_{\mu \nu} 
\hat{W}^{\mu \nu} \Phi  \; , \nonumber \\
{\cal O}_{BB} & = & \Phi^{\dagger} \hat{B}_{\mu \nu} 
\hat{B}^{\mu \nu} \Phi \; ,  \nonumber \\
{\cal O}_W  & = & (D_{\mu} \Phi)^{\dagger} 
\hat{W}^{\mu \nu}  (D_{\nu} \Phi) \; , \nonumber \\
{\cal O}_B  & = &  (D_{\mu} \Phi)^{\dagger} 
\hat{B}^{\mu \nu}  (D_{\nu} \Phi)  \; , 
\label{eff} 
\end{eqnarray}
where $\Phi$ is the Higgs doublet in the unitary gauge $\Phi = (v
+ H)/\sqrt{2} \; (0 \;\; 1)^T$, $\hat{B}_{\mu \nu} = i (g'/2)
B_{\mu \nu}$, and $\hat{W}_{\mu \nu} = i (g/2) \sigma^a W^a_{\mu
\nu}$,  with $B_{\mu \nu}$ and $ W^a_{\mu \nu}$ being the field
strength tensors of the respective $U(1)$ and $SU(2)$ gauge
fields, and $D_\mu = \partial_\mu + i g T^a  W^a_\mu + i g' Y
B_\mu$, the covariant derivative. 

The anomalous $H\gamma\gamma$ and $HZ\gamma$ couplings generated
by (\ref{eff}), can be written in a compact form as,
\begin{equation}
{\cal L}_{\text{eff}}^{\text{H}} = 
g_{H \gamma \gamma} H A_{\mu \nu} A^{\mu \nu} + 
g^{(1)}_{H Z \gamma} A_{\mu \nu} Z^{\mu} \partial^{\nu} H + 
g^{(2)}_{H Z \gamma} H A_{\mu \nu} Z^{\mu \nu}
\; , 
\label{H} 
\end{equation} 
where $A(Z)_{\mu \nu} = \partial_\mu A(Z)_\nu - \partial_\nu
A(Z)_\mu$, and  the coupling constants $g_{H \gamma \gamma}$, and
$g^{(1,2)}_{H Z \gamma}$ are related to the coefficients of the
operators appearing in (\ref{l:eff}) through, 
\begin{eqnarray}
g_{H \gamma \gamma} &=& - \left( \frac{g M_W}{\Lambda^2} \right)
                       \frac{s^2 (f_{BB} + f_{WW} - f_{BW})}{2} \; , 
\nonumber \\
g^{(1)}_{H Z \gamma} &=& \left( \frac{g M_W}{\Lambda^2} \right) 
                     \frac{s (f_W - f_B) }{2 c} \; ,  \\
\label{g}
g^{(2)}_{H Z \gamma} &=& \left( \frac{g M_W}{\Lambda^2} \right) 
                      \frac{s (2 s^2 f_{BB} - 2 c^2 f_{WW} + 
                     (c^2-s^2)f_{BW} ) }{2 c} \nonumber \; , 
\end{eqnarray}
with $g$ being the electroweak coupling constant, and $s(c)
\equiv \sin(\cos)\theta_W$. It is important to point out that the
operators (\ref{eff}) does not induce any other new  tree--level
anomalous contributions besides the ones leading to the
$H\gamma\gamma$ and $HZ\gamma$ interactions. In particular,
$4$--point anomalous couplings like $Z \gamma \gamma \gamma$ and
$\gamma \gamma \gamma \gamma$ are absent.

In order to reduce the number of free parameters and, at the same
time, relate the anomalous Higgs and the triple vector boson
couplings, we make the natural assumption that all the
coefficients of the dimension 6 operators have a common value $f$
\cite{dpf}. In this scenario, $g^{(1)}_{H Z \gamma} = 0$, and we
can relate the other Higgs boson anomalous couplings with  the
coefficient of the usual anomalous vector boson coupling, $\Delta
\kappa_\gamma = (M_W^2/\Lambda^2) f$, through,
\begin{eqnarray}
|g_{H \gamma \gamma}| & = & \frac{g s^2}{2 M_W} 
                            |\Delta \kappa_\gamma| 
\simeq 9.2 \times 10^{-4} \; \text{GeV}^{-1} \times 
|\Delta \kappa_\gamma|,
\nonumber \\ 
|g^{(2)}_{H Z\gamma}| & = & \frac{g s c (1 - s^2/c^2)}{2 M_W} 
                            |\Delta \kappa_\gamma|
\simeq 1.2 \times 10^{-3} \; \text{GeV}^{-1} \times 
|\Delta \kappa_\gamma|.
\label{ghgg}
\end{eqnarray}

We should point out that the anomalous vector boson couplings,
like $\Delta \kappa_\gamma$, are basically unconstrainted by the
current high precision electroweak data. Presently, the best
direct bound on $\Delta \kappa_\gamma$ was the one obtained by
the CDF Collaboration \cite{cdf} and constraints  $-1.0 < \Delta
\kappa_\gamma < 1.1$, at 95\% C.L.. At LEP2, the angular
distribution of final state fermions of the reaction $e^+ e^- \to
W^+ W^- \to \ell \nu j j$ will be able to  further restrict the
allowed values of $\Delta \kappa_\gamma$ to $-0.19 < \Delta
\kappa_\gamma < 0.21$, at 95\% C.L. \cite{dpf}, for $\sqrt{s} =
176$ GeV. 

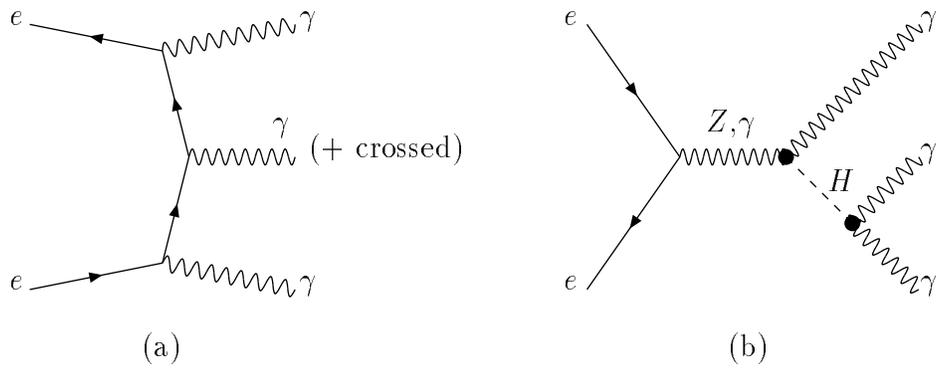
\begin{figure}
\protect
\caption{Feynman diagrams for  $e^+ e^- \to \gamma\gamma\gamma$}
\label{fig:1}
\end{figure}

An interesting option to test the couplings described by
(\ref{H}) is through the reaction $e^+ e^- \to \gamma, \; Z \to
\gamma H (\to \gamma \gamma)$. In the SM, the process  $e^+ e^-
\to \gamma\gamma\gamma$, at tree--level, is purely
electromagnetic and is represented by the diagrams of Fig.\
\ref{fig:1}(a). The inclusion of anomalous Higgs boson couplings
gives rise to two additional contributions presented in Fig.\
\ref{fig:1}(b). Since we are assuming that the Higgs boson
coupling with fermions are the standard ones, {\it i.e.\/}
proportional to $m_f/v$, we have neglected the corresponding
contributions to this process.

This process has already been tested at the Z pole by LEP1
\cite{exp} which  established an upper limit on the branching
ratio $B(Z \to \gamma \gamma \gamma) < 1.0 \times 10^{-5}$. We
found that this bound is not able to further restrict
$\Delta\kappa_{\gamma}$ beyond the existent direct bounds
\cite{cdf}. For instance, for a 70 GeV Higgs, the above limit
requires $\Delta\kappa_{\gamma} > 1.4$.

Here we make a detailed analysis for the two expected runs of
LEP2 collider {\it i.e.\/}  $\sqrt{s} = 176$ GeV, with  ${\cal L}
= 0.5$ fb$^{-1}$, and  $\sqrt{s} = 190$ GeV, with  ${\cal L} =
0.3$ fb$^{-1}$. We performed a Monte Carlo analysis using  the
package MadGraph \cite{Madgraph} coupled to HELAS \cite{helas}.
Special subroutines were constructed for the anomalous
contributions which enable us to take into account all
interference effects between the QED and the anomalous
amplitudes. The phase space integration was performed by VEGAS
\cite{vegas}. 

In order to search for optimum cuts to maximize the signal to
background ratio, we label the three final state photons as
$\gamma_{1,2,3}$ according to decreasing value of their energy,
{\it i.e.\/}  $E_{\gamma_1} > E_{\gamma_2} > E_{\gamma_3}$.
We start our analysis applying the following standard cuts
on the events \cite{l3}, 
\begin{eqnarray}
| \cos \theta_{e i} | & \leq & 0.97 \; , \\
\theta_{i j} & > & 15^\circ \; , \\
E_{\gamma_i} & \geq & 5 \; \text{GeV} \; . 
\label{cut:1} 
\end{eqnarray} 
where $\theta_{e i}$ is the angle between the photon $i$ and the
electron/positron beam direction, and $\theta_{i j}$, is the
angle between the photon pair $(i, j)$.

In Figures 2--4, we present our results for the photon spectra
and angular  distributions where only these cuts were applied.
For illustrative purposes we consider a center--of--mass energy
of $\sqrt{s} = 176$ GeV,  $M_H = 80$  GeV and an anomalous
coupling $\Delta \kappa_\gamma$ =  1.5. 

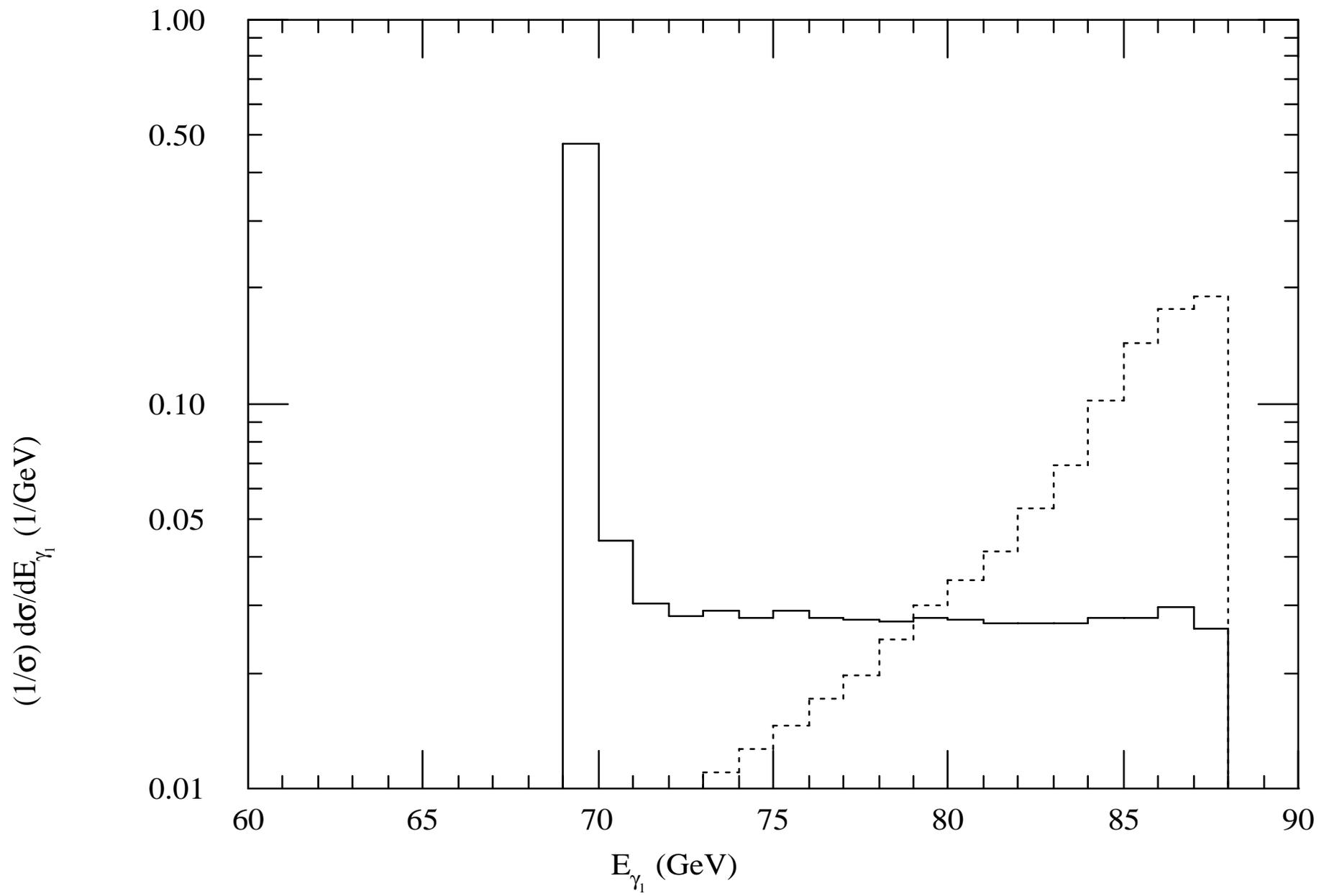
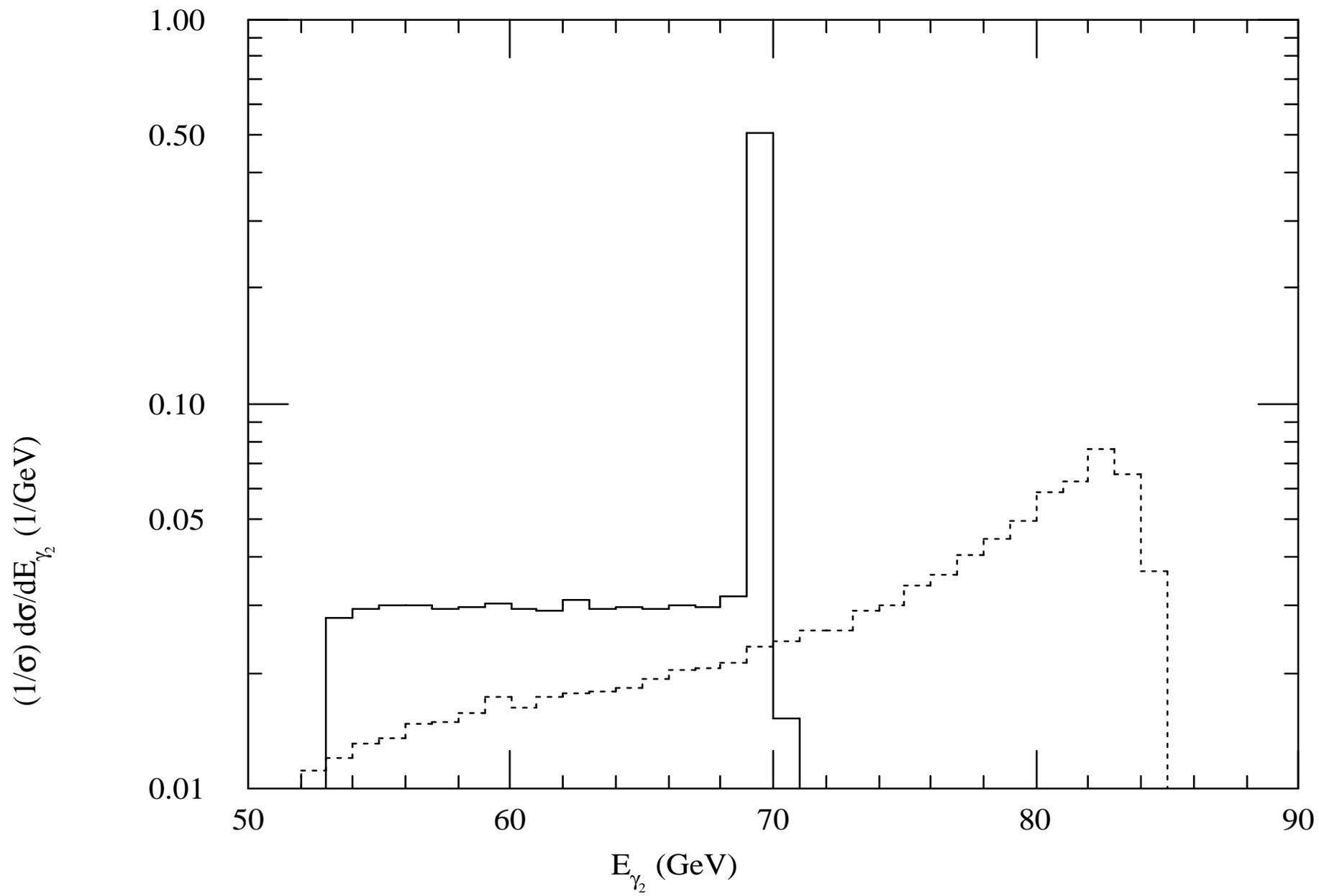
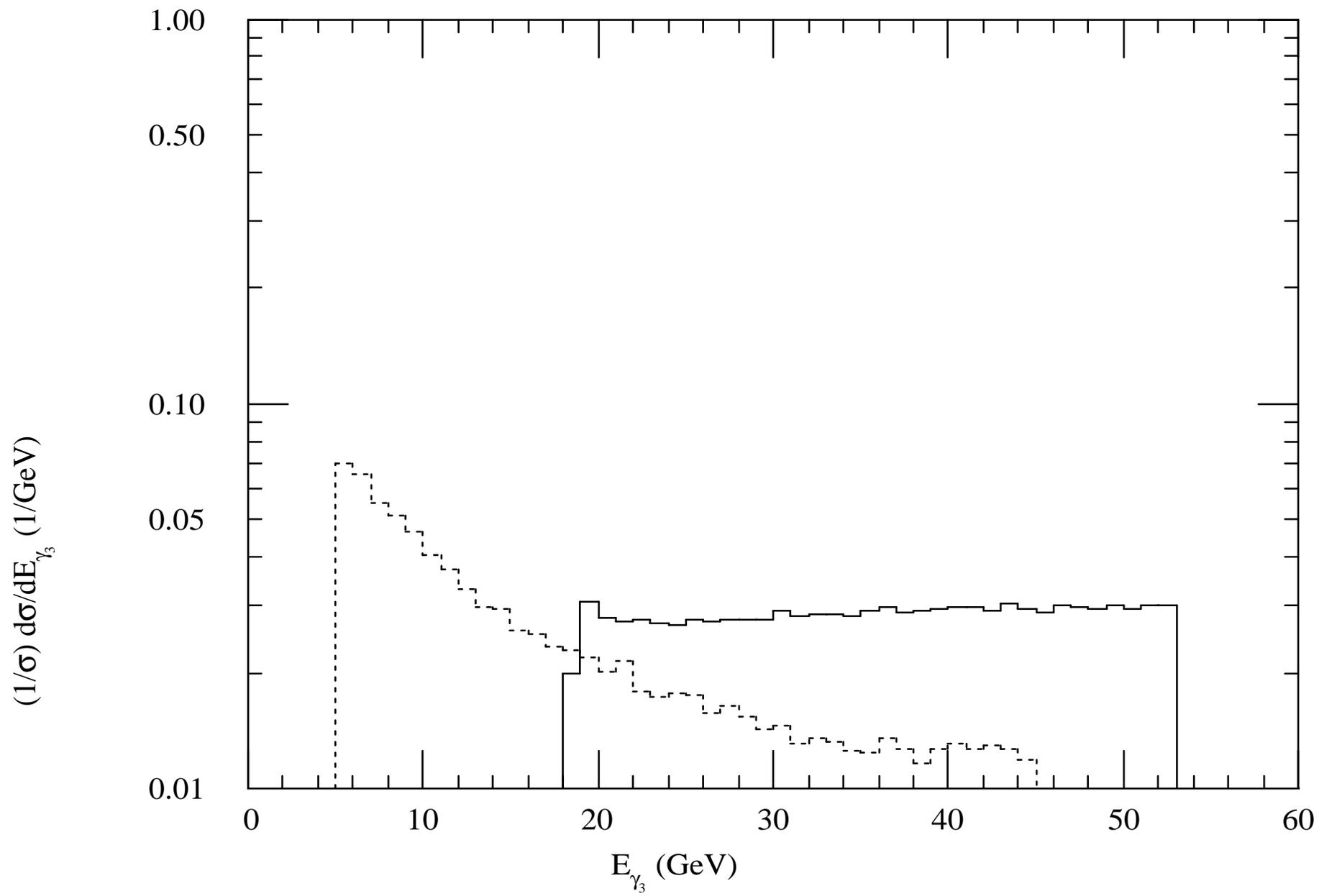
\begin{figure}
\protect
\caption{Normalized photon energy distribution
$(1/\sigma)d\sigma/dE_{\gamma_i}$ ($i = 1, 2, 3$), for the
anomalous  (solid line) and standard model (dashed line)
contributions. We assumed $\Delta \kappa_\gamma =  1.5$ and $M_H
= 80$ GeV.}
\label{fig:2}
\end{figure}

In Fig.\ \ref{fig:2}, we compare the normalized photon energy
spectra of the signal and background, for the three photons. We
can notice that for the background the energy of the two most
energetic photons, $E_{\gamma_{1},\gamma_{2}}$, tends to be close
to $\sqrt{s}/2$, whereas the least energetic one, which is
emitted by bremsstrahlung, is very soft. The signal is dominated
by on--mass--shell $H \gamma$ production, with the subsequent
decay $H \to \gamma \gamma$. The photon that does {\it not\/}
come from the Higgs boson decay tends to have a two--body
spectrum with energy $E_{\gamma} = (s - M_H^2)/(2 \sqrt{s})$.
This behavior is evident in Fig.\ \ref{fig:2} (a) and (b) where
the peak at $E_{\gamma} \simeq 69.8$ GeV, is due to this
``monochromatic" component. On the other hand,  the softest
photon, that always come from the Higgs boson decay, has an
minimum energy given by  $E_{\gamma_3}^{\text{min}} \simeq
M_H^2/(2\sqrt{s})$.

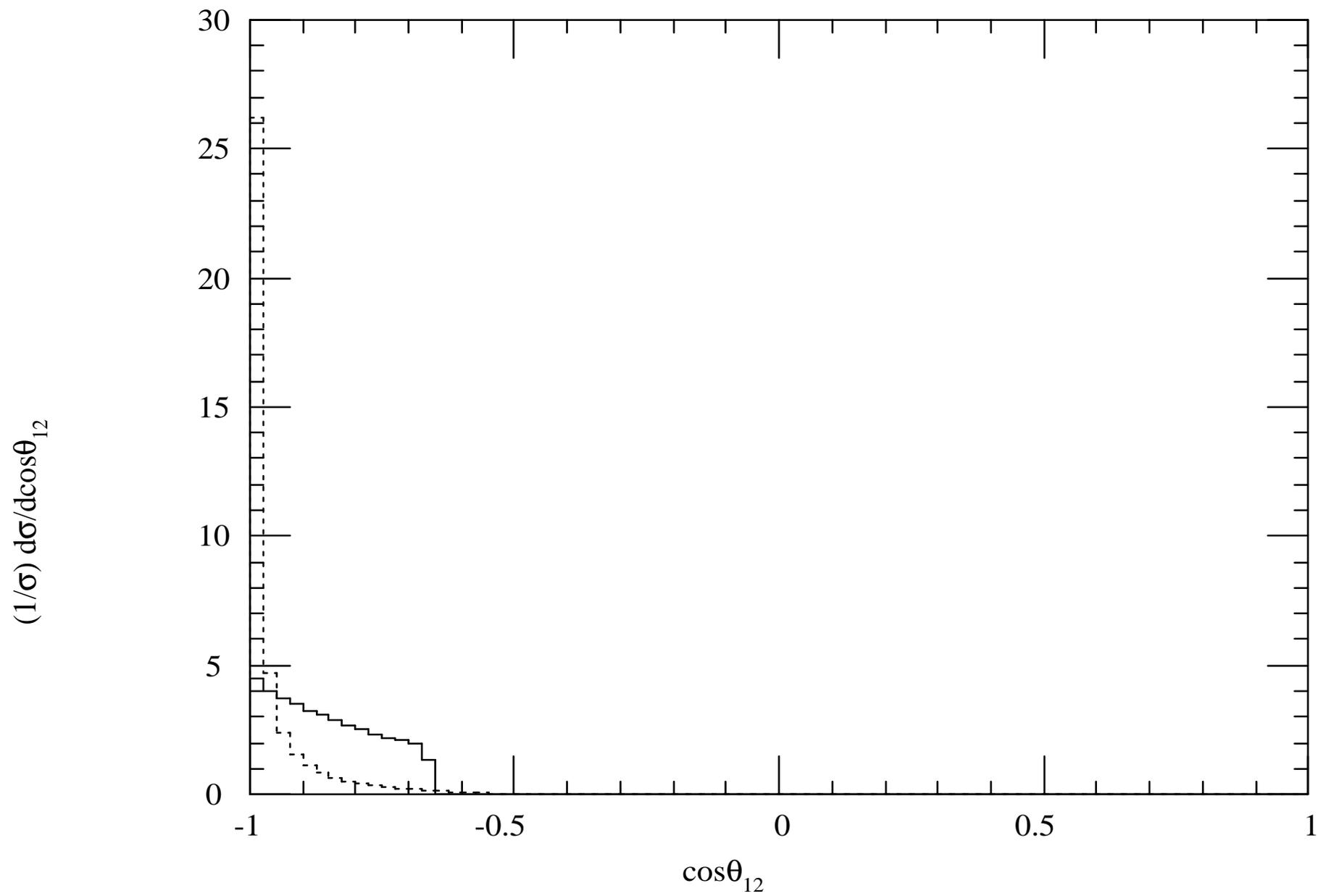
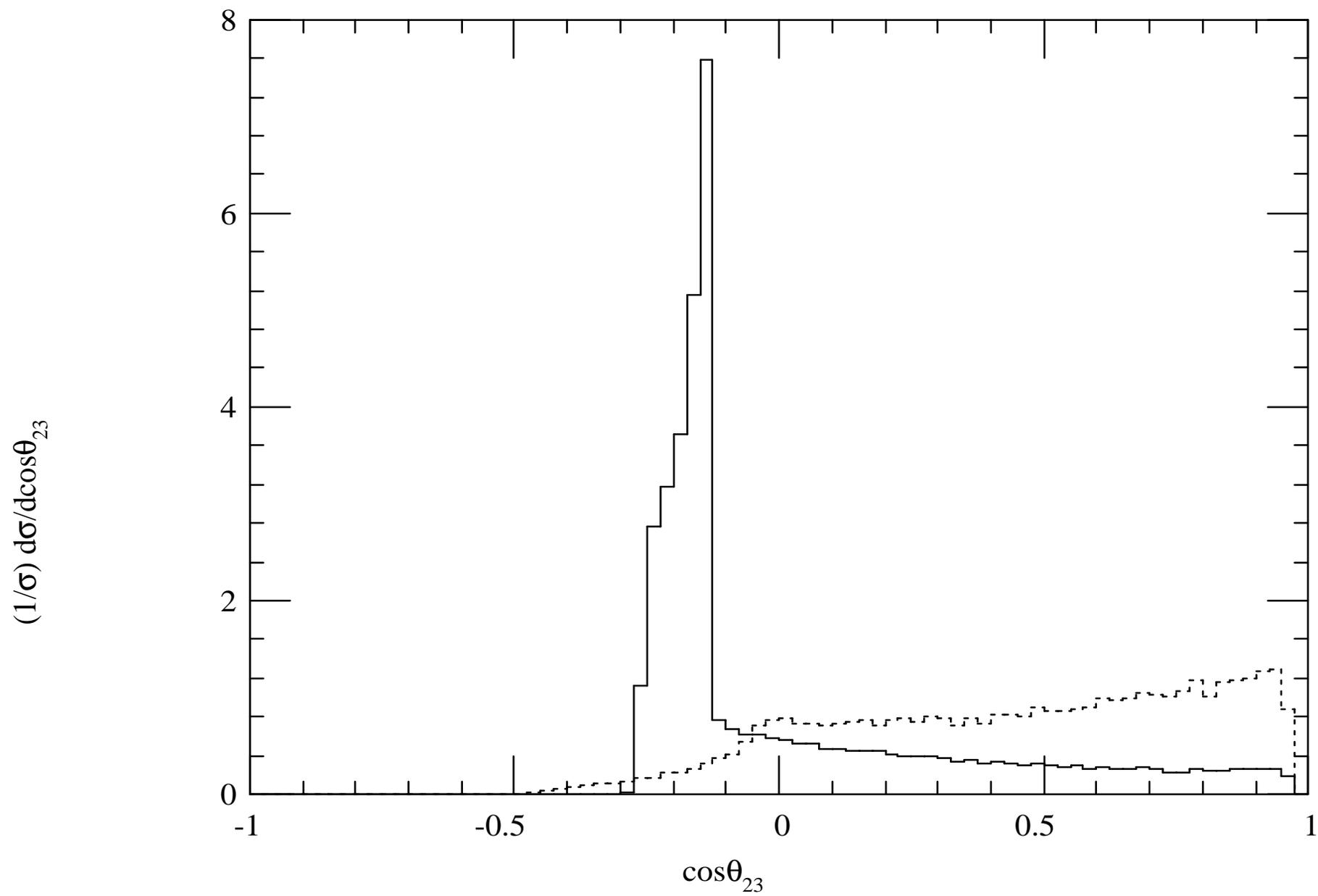
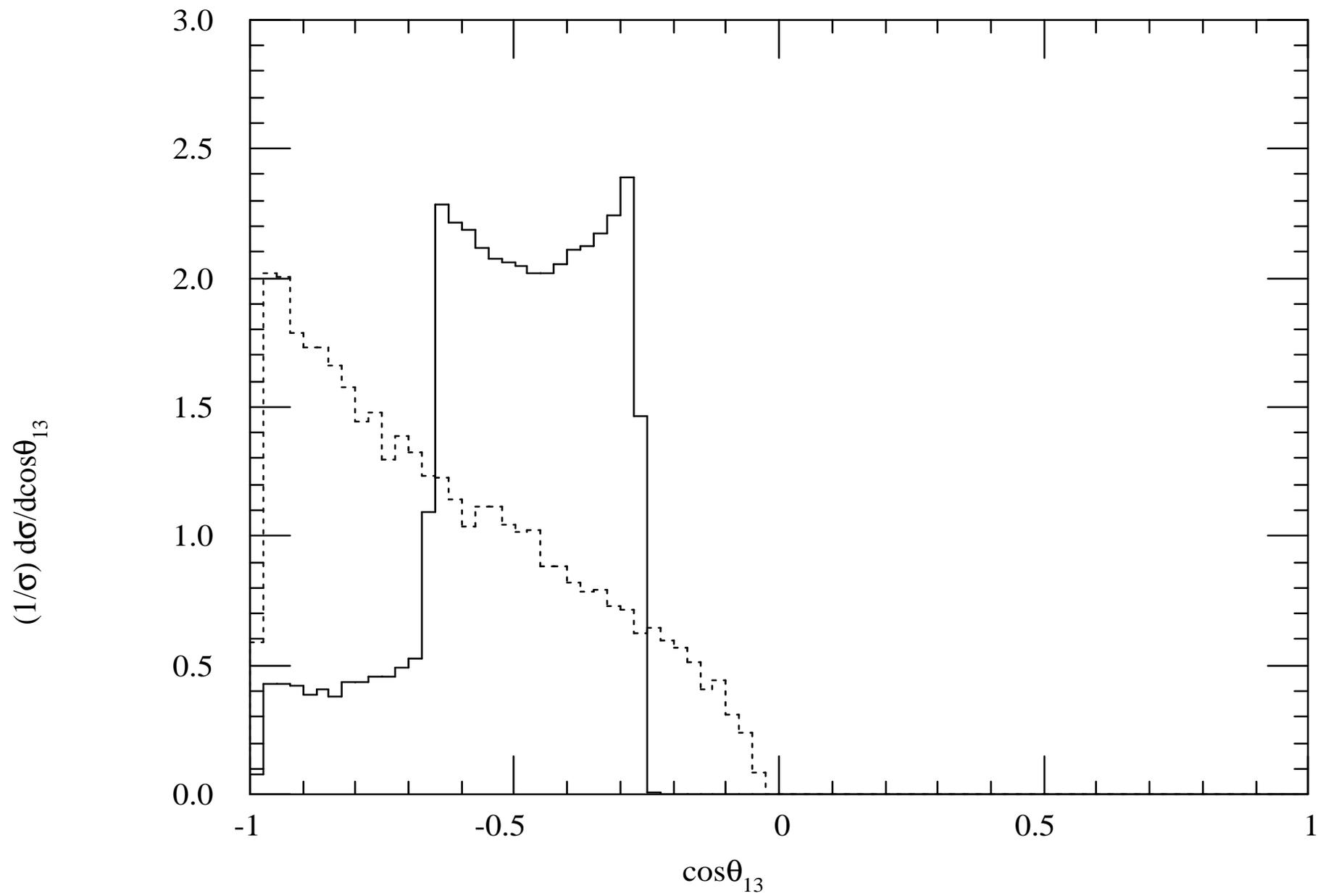
\begin{figure}
\protect
\caption{Normalized angular distribution between photons $(1/\sigma)
d\sigma/d\cos\theta_{ij}$. The conventions are the same as in Fig.\
\protect{\ref{fig:2}}.}
\label{fig:3}
\end{figure}

In Fig.\ \ref{fig:3}, we present the normalized  distribution of
all photon pair angles. We can learn from Fig.\ \ref{fig:3} (a)
that for the background the two most energetic photons are almost
back--to--back, while for the signal this distribution is
broader.  The angular distributions for $\theta_{23}$ and
$\theta_{13}$ reflect the existence of a minimum attainable angle
between the two photons coming from the Higgs boson,
$\theta^{\text{min}}_{\gamma\gamma} (H) =  2 \; \arcsin
(M_H/E_H)$.

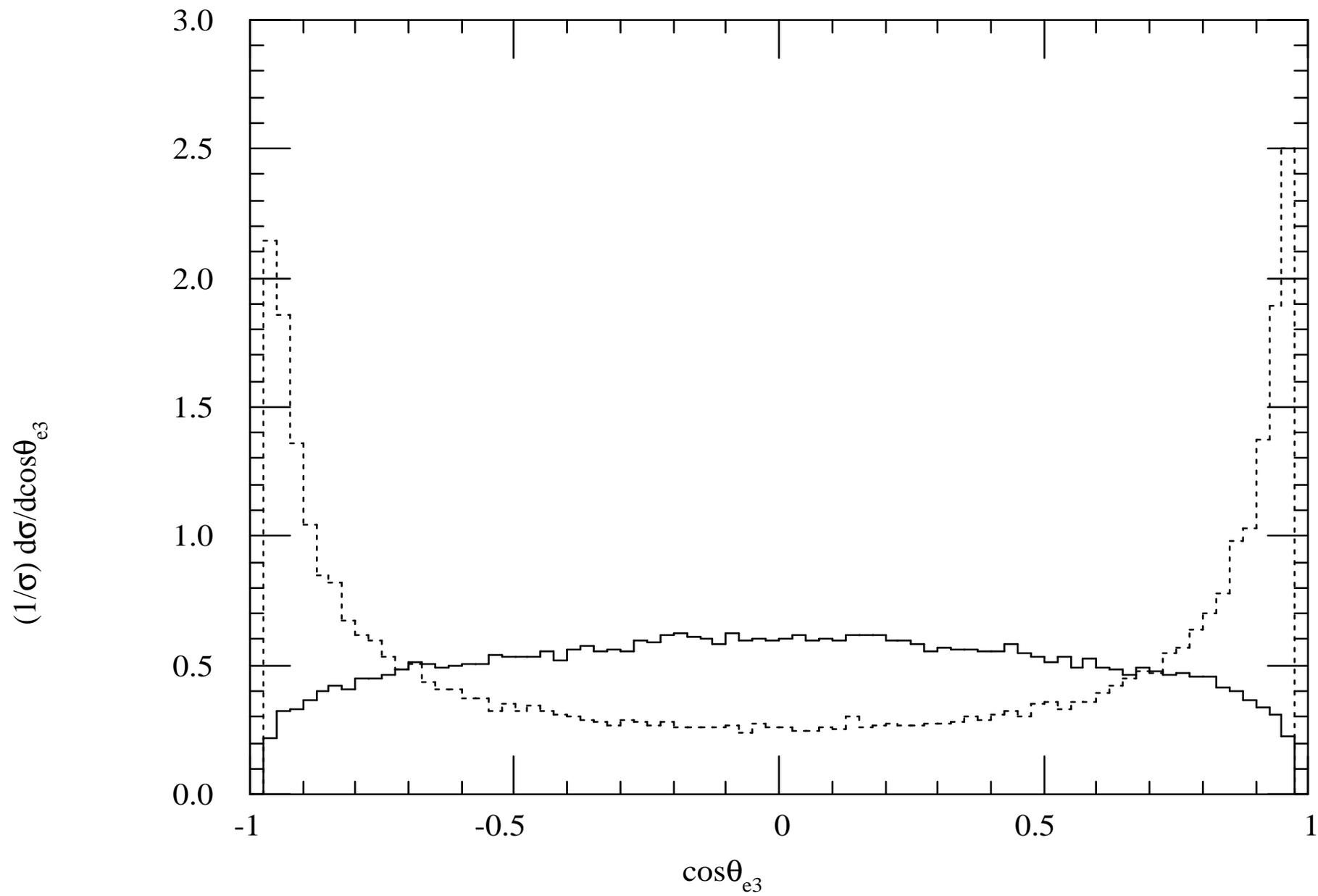
\begin{figure}
\protect
\caption{Normalized angular distribution of the softest photon
with the beam $(1/\sigma) d\sigma/d\cos\theta_{e3}$. The
conventions are the same as in Fig.\ \protect{\ref{fig:2}}.}
\label{fig:4}
\end{figure}

In Fig.\ \ref{fig:4}, we present the normalized angular
distribution between the least energetic photon, $\gamma_3$, and
the electron beam. This distribution, for the background,  is
peaked in the forward and backward directions, while for the
signal the least energetic photon, that always come from the
Higgs boson decay, has a flat distribution with the beam. We omit
the angular distributions between the other two photons with the
beam since the behavior of the signal and background are quite
similar.

An analysis of the angular distributions (Fig.\ \ref{fig:3})
suggests that, if we require the maximum angle between any pair
of photons to be less than $165^\circ$ ($\cos \theta_{ij} >
-0.97$), we can eliminate a large portion of the background
events. When this cut is implemented the background falls from
$699$ fb  to $203$ fb while the signal remains almost the same,
going from $36.1$ to $30.7$ fb. On the other hand, from the
photon  spectra  (Fig.\ \ref{fig:2}) we learn that cutting the
maximum energy of each photon, for instance $E_\gamma < 70$ GeV,
can significantly reduce the background. When we introduce this
cut the total cross section of the background falls $\sim 90 \%$,
while the signal is reduced by less than one half. Finally,  the
angular distribution of the softest photon with the beam suggests
a  cut $|\cos \theta_{e3}| \leq 0.8$, which reduces the
background to the same order of the signal cross section. We
should notice that the above cuts are less efficient for heavier
Higgs bosons, restricting our analysis to a Higgs boson with mass
up to 100 GeV.

From the above considerations, we have further imposed the energy cut,
\begin{eqnarray}
5 \leq E_{\gamma_{i}}  & \leq & 70 \; (80) \; \mbox{GeV}   \;, 
\end{eqnarray} 
for $\sqrt{s} = 176$ $(190)$ GeV, and the following angular cuts,
\begin{eqnarray} 
| \cos \theta_{e 1, e 2} | & \leq & 0.97 \; , \\
| \cos \theta_{e3} | &  \leq & 0.80 \; , \\
15^\circ < \theta_{i j} & < & 165^\circ \; .
\end{eqnarray} 

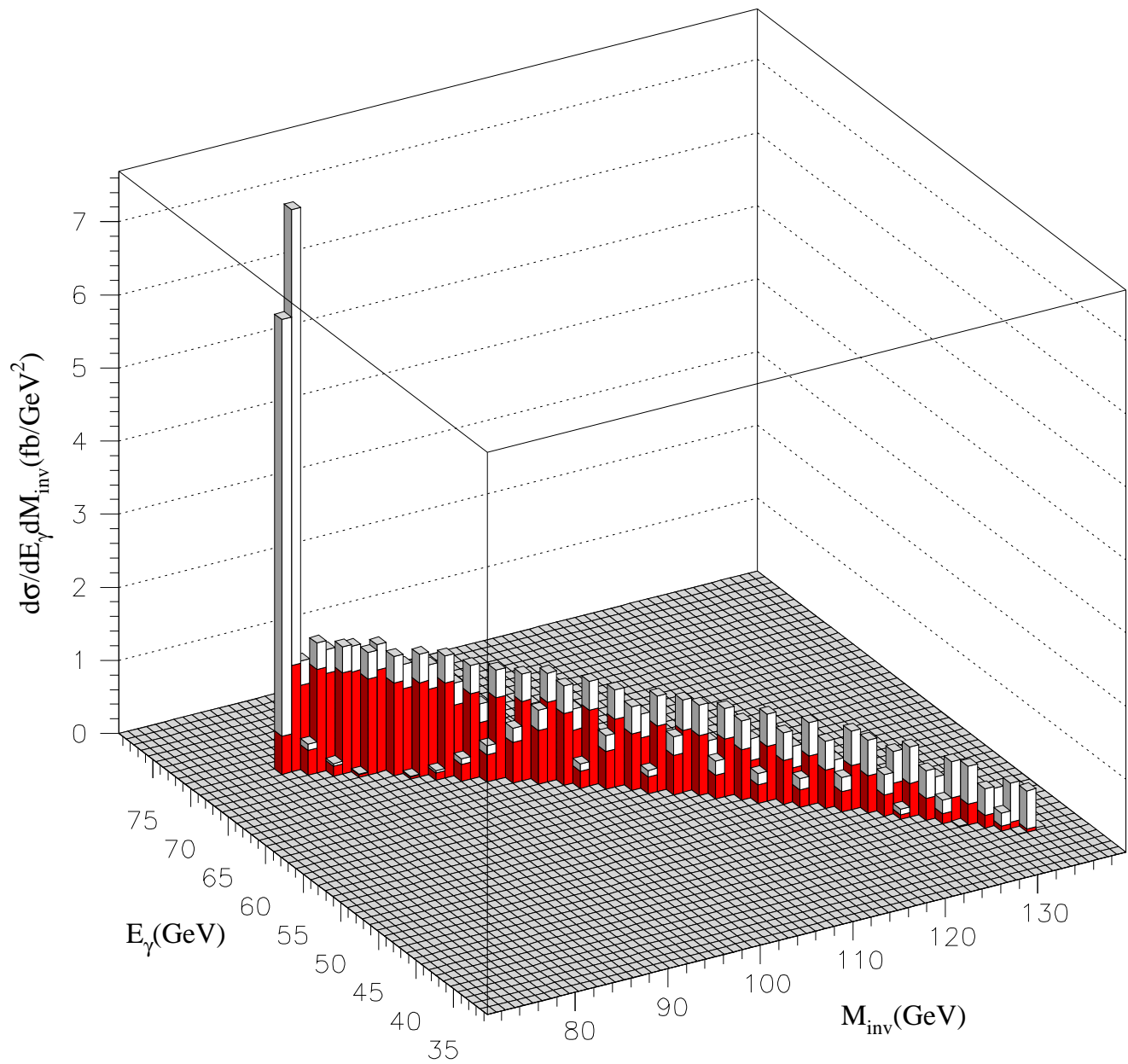
\begin{figure}
\protect
\caption{Double photon energy--invariant mass distribution
$d\sigma/dE_{\gamma}dM_{inv(\gamma)}$,  for the process $e^+ e^-
\to \gamma\gamma\gamma$, for the background (shaded histogram)
and for the signal (white histogram) with $\Delta \kappa_\gamma
=  1.5$ and $M_H = 80$ GeV.}
\label{fig:5}
\end{figure}

In Fig.\ \ref{fig:5}, we present the double differential
distribution of the energy and invariant mass,
$d\sigma/dE_{\gamma} dM_{\text{inv}}$, for $\Delta \kappa_\gamma
= 1.5$ and $M_H = 80$ GeV.  We have added up all events with
$E_{\gamma_i}$, and $M_{\text{inv}_{jk}}$, for $i \neq j \neq k$.
This Figure shows very clearly the enhancement  due to anomalous
Higgs contribution around $M_{\text{inv}} \simeq 80$ GeV and
$E_\gamma \simeq 70$ GeV.

\section{Conclusions}
\label{sec:con}

In order to estimate the reach of LEP2 to disentangle the
anomalous Higgs boson couplings via the reaction $e^+ e^- \to
\gamma \gamma \gamma$, we have evaluated the significance of the
signal based both on the total cross section and on the Higgs
boson enhancement in the $d\sigma/dE_{\gamma} dM_{\text{inv}}$
distribution. We have scanned the values of  $\Delta
\kappa_\gamma$, for $M_H =$ 80, 90, and 100 GeV. 

We show in Table \ref{tab:1} the minimum values of $\Delta
\kappa_\gamma$ that can be probed in two runs of LEP2, for a
center--of--mass energy of $176$ and $190$ GeV, with luminosities
of  ${\cal L} = 0.5$ and  $0.3$ fb$^{-1}$, respectively. The
combined result for both runs is also presented. We required
a 95 \% of C.L.\ effect in the total cross section
($\sigma_{\text{tot}}$) and also in the double differential
distribution ($d\sigma/dE_{\gamma} dM_{\text{inv}}$). In the
latter case, we have added up the events in the eight 1 GeV bins
around the expected Higgs boson signal. 

We found that, if the anomalous coupling  $|\Delta \kappa_\gamma|
\gtrsim 0.8$ it will be possible to identify an anomalous Higgs
boson in the range 80--100 GeV with 95 \% C.L.. However, the
signature of a heavier Higgs boson will not be so clear since the
reaction $e^+ e^- \to \gamma, \; Z \to \gamma H (\to \gamma
\gamma)$ is particularly important when the Higgs boson is almost
on--mass shell.

In conclusion, the search for anomalous Higgs boson couplings
at LEP2 provides a complementary way to probe effective
Lagrangians that are low--energy limit of physics beyond the SM.
We have shown that the study of the process  $e^+ e^- \to \gamma
\gamma \gamma$ is able to improve the present limits on the
anomalous vector boson couplings that are concomitantly involved
in operators like (\ref{eff}) that modify the bosonic sector of the
Standard Model.  

\acknowledgments

This work was supported by Conselho Nacional de Desenvolvimento
Cient\'{\i}fico e Tecnol\'ogico (CNPq), and by Coordena\c{c}\~ao
de Aperfei\c{c}oamento de Pessoal de N\'{\i}vel Superior (CAPES).

\widetext
\begin{table}
\begin{tabular}{||c||c|c||c|c||c|c||}
& \multicolumn{2}{c||}{$\sqrt{s} = 176$ GeV} & 
\multicolumn{2}{c||}{$\sqrt{s} = 190$ GeV} &
\multicolumn{2}{c||}{Combined} \\
\hline 
\hline 
$M_H$ (GeV) & $\sigma_{\text{tot}}$ & $d\sigma/dE_{\gamma} dM_{\text{inv}}$ & 
$\sigma_{\text{tot}}$ & $d\sigma/dE_{\gamma} dM_{\text{inv}}$ &
$\sigma_{\text{tot}}$ & $d\sigma/dE_{\gamma} dM_{\text{inv}}$ \\  
\hline 
\hline 
80 & 1.32 & 0.89 & 1.49 & 0.96 & 1.18 & 0.78 \\
\hline 
90 & 1.42 & 0.95 & 1.56 & 0.97 & 1.26 & 0.81 \\
\hline 
100 & 1.64 & 1.04 & 1.70 & 1.01 & 1.41 & 0.87 \\
\end{tabular}
\caption{Values of $\protect{\Delta \kappa_{\gamma}}$,
corresponding to 95 \% C.L. from the total cross section
$\protect{\sigma_{\text{tot}}}$, and from the double distribution
$\protect{d\sigma/dE_{\gamma} dM_{\text{inv}}}$, for the 
$\protect{\sqrt{s} = 176}$ and 190 GeV runs of LEP2.}
\label{tab:1}
\end{table}


\begin{references}

\bibitem{hep:conf} The LEP Collaborations ALEPH, DELPHI, L3,
OPAL, and the LEP Electroweak Working Group, contributions to the
1995 Europhysics Conference  on High Energy Physics
(EPS--HEP), Brussels, Belgium and to the 17th International
Symposium on Lepton-Photon Interactions,  Beijing, China,
Report No. CERN-PPE/95-172 (1995).

\bibitem{effective} S.\ Weinberg, Physica {\bf 96A} (1979) 327;
J.\ F.\ Donoghue, E.\ Golowich and B.\ R.\ Holstein, {\it
Dynamics of the Standard Model} (Cambridge University Press,
Cambridge, England, 1992). 

\bibitem{Hagiwara2} K.\ Hagiwara, R.\ Szalapski, and D.\
Zeppenfeld, Phys.\ Lett.\ {\bf B318} (1993) 155.

\bibitem{epem} K.\ Hagiwara, and M.\ L.\ Stong, Z.\ Phys.\ {\bf
C62} (1994) 99; 
B.\ Grzadkowski, and J.\ Wudka, Phys.\ Lett.\ {\bf B364} (1995) 49; 
G.\ J.\ Gounaris, J.\ Layssac and F.\ M.\ Renard, Z.\ Phys.\ {\bf
C65} (1995) 245; 
G.\ J.\ Gounaris, F.\ M.\ Renard and N.\ D.\ Vlachos, Nucl.\
Phys.\ {\bf B459} (1996) 51; W.\ Kilian, M.\ Kr\"amer and P.\
M.\ Zerwas, Report No.\ hep--ph/9603409;  
S.\ M.\ Lietti, S.\ F.\ Novaes and R.\ Rosenfeld, Phys.\ Rev.\ D, 
in press.

\bibitem{gg} G.\ J.\ Gounaris and F.\ M.\ Renard, Z.\ Phys.\ {\bf
C69} (1996) 513.

\bibitem{3gamma} See, for instance: M. \ Baillargeon, F. \
Boudjema, E. \ Chopin and V. \ Lafage, Report No.\
hep--ph/9506396 and references therein.

\bibitem{dim6:zep} K.\ Hagiwara, S.\ Ishihara, R.\ Szalapski and
D.\ Zeppenfeld, Phys.\ Rev.\ D {\bf 48} (1993) 2182.

\bibitem{dpf} H.\ Aihara {\it et al.}, in ``Electroweak Symmetry
Breaking and Beyond the Standard Model", edited by T.\ Barklow,
S.\ Dawson, H.\ Haber and J.\ Siegrist (World Scientific, Singapore),
 Report No.\ hep--ph/9503425.

\bibitem{cdf} S.\ M.\ Errede, in Proceedings of the XXVII
International Conference on High Energy Physics,  Glasgow,
Scotland, p.\ 433, edited by P.\ J.\ Bussey and and I.\ G.\
Knowles. 

\bibitem{exp}OPAL Collaboration, Phys.\ Lett.\ {B257} (1991) 531;
ALEPH Collaboration, Phys.\ Rep.\ {\bf 216} (1992) 253;
L3 Collaboration, Phys.\ Lett.\ {\bf B353} (1995)
136; {\it idem} Phys.\ Lett.\ {\bf B345} (1995) 609;
DELPHI Collaboration, Phys.\ Lett.\ {\bf B327} (1994) 386;

\bibitem{Madgraph} T.\ Stelzer and W.\ F.\ Long, Comput.\
Phys.\ Commun.\ {\bf 81} (1994) 357. 

\bibitem{helas} H.\ Murayama, I.\ Watanabe and K.\ Hagiwara,
KEK Report 91-11 (unpublished).

\bibitem{vegas} G.\ P.\ Lepage,  J. \ Comp. \ Phys. {\bf 27}
(1978) 192; ``Vegas: An Adaptive Multidimensional Integration
Program", CLNS-80/447, 1980 (unpublished).

\bibitem{l3} L3 Collaboration, Phys.\ Lett.\ {\bf B353} (1995)
136.

\end{references}
\end{document}